\documentstyle[12pt]{article}
\begin{document}
\def\strut{\rule[-.5cm]{0cm}{1cm}}
\def\dspace{\baselineskip = .30in}

\title{
\begin{flushright}
{\large \bf IC/94/156}\\
{\large\bf IFUP-TH 39/94}
\end{flushright}
\vspace{1.5cm}
\Large\bf Semiglobal Alice Strings }

\author{{\bf G.  Dvali} \thanks{Permanent address: Institute of
Physics, Georgian Academy of Sciences, 380077 Tbilisi, Georgia}\\
Dipartimento di Fisica, Universita di Pisa and INFN,\\
Sezione di Pisa I-56100 Pisa, Italy\\E-mail:
dvali@mvxpi1.difi.unipi.it
\and
{\bf Goran Senjanovi{\'c}}\\
International Centre for Theoretical Physics\\
Trieste, Italy\\ E-Mail: goran@ictp.trieste.it}

\date{ }
\maketitle

\begin{abstract}
\bf We show that in certain  theories with topologically trivial quotient space
of spontaneously broken gauge symmetry there  can exist
topologically stable strings  that carry nonabelian gauge flux. These objects
result from the ``accidental'' global degeneracy of the vacuum which makes it
topologically nontrivial.
 In particular, some models contain the semiglobal analog of the Alice
strings.
\end{abstract}
\newpage

\dspace

{\bf Introduction}. Many particle physics  models  in the  modern cosmological
context
lead to the formation of  stable
vacuum defects. Their origin has to do  with  the topology of the vacuum
manifold $V$. When the first homotopy group $\pi_1 (V)$ is
nontrivial, the resulting defects are strings.
If $V$ is topologically equivalent to the quotient space $G_l/H_l$ of the
spontaneously broken gauge symmetry $G_l \rightarrow H_l$ (meaning that
there is no additional global degeneracy in $V$), strings carry a topologically
stable gauge flux which compensates the gradient of the Higgs field at
spatial infinity \cite{N-O}. Such strings are called  local (or gauge) strings.

Due to the above, it is believed sometimes that the existence of
topologically stable flux carrying strings can be recognised by examining
the first homotopy group of the quotient space $G_l/H_l$. That is, the flux
persists (topologically) only if $\pi_1(G_l/H_l)\neq 1$. This criterion
is based on the assumption that even if the actual global structure of $V$
is larger than $G_l/H_l$ (so that $\pi_1(V)\neq 1$), at best the resulting
defect can be global strings empty of flux. As we have shown recently
\cite{semiglobal}
this assumption is false: topologically stable flux carrying strings can
exist even if $\pi_1 (G_l/H_l) = 1$, provided the global vacuum manifold
$V$ is topologically nontrivial. That is

\begin{equation}
\pi_1 (G_{gl}/H_{gl}) \neq 1
\end{equation}

 where $G_{gl}$ (covering $G_l$) and $H_{gl}$ (covering $H_l$) are
the initial and final global symmetry groups respectively. In particular, as
was shown in \cite{semiglobal}, such topologically stable flux carrying strings
 exist  automatically in the    $SU(2)\otimes U(1)$ electroweak model
 provided the theory contains
  at least one spontaneously broken global $U(1)_{gl}$ factor under which the
electroweak Higgs doublet(s) transforms nontrivialy. In particular the role
of $U(1)_{gl}$ can be played by $R$-symmetry, Peccei-Quinn  or $(B-L)$
symmetry.
The topologically stable flux carried by such strings is rather unusual: it
does not compensate the logarithmic divergence of the Higgs gradient energy
at  spatial infinity.    Therefore strings carry  properties of both
global and gauge strings, and in \cite{semiglobal} we have called them
``semiglobal''.
The semiglobal strings should not be confused with  the semilocal ones
\cite{sls} for
which the situation is just opposite:
$\pi_1 (G_l/H_l) \neq 1$  whereas
$\pi_1 (G_{gl}/H_{gl} ) = 1$, and the flux is always topologically
unstable (while it can be classically stable in  a certain range of
parameters).
Semiglobal strings discussed in \cite{semiglobal} are of Abelian nature, since
their
stability is due to the breaking of an  Abelian $U(1)_{gl}$ factor that renders
$V$ topologically nontrivial.

 In the present paper we investigate the existence of   nonabelian
semiglobal strings. In particular we show that  certain realistic models
contain the semiglobal analogs of ``Alice'' strings \cite{alice}. This strings
are
formed due to the breaking of ``accidental'' nonabelian global symmetries
of the classical Higgs potential which satisfy (1). These symmetries are
accidental in the sence that they are not respected by all (and in
particular by gauge) interactions of the theory, but are automatically
compatible with {\em all possible} gauge invariant terms of the classical
Higgs potential. Therefore at the classical level they provide  a
topological nontrivial degeneracy of the vacuum leading to the formation
of  strings.   The quantum corrections in
general lift the accidental nonabelian degeneracy of the vacuum and make it
topologically trivial. However the strength of this correction is
controlled by the gauge coupling constant so that they do not alter the
classical stability of the flux in
a wide range of parameters. The above consideration provides a crucial
difference between
abelian and nonabelian semiglobal strings: in contrast with nonabelian ones,
the accidental abelian symmetries of the scalar potential are automatically
respected by  the gauge interaction and appear to be true global
symmetries of
the theory (unless broken by anomalies or by Yukawa interactions).
Consequently, the abelian semiglobal strings remain topologically stable
even after quantum corrections.

\dspace

{\bf Topologically stable nonabelian flux}.
 As an example consider the simplest model that gives rise to a gauge Alice
string \cite{alice}. Such is the one with an $SO(3)$-local symmetry
spontaneously
broken down to  the semidirect product $U(1) \otimes Z_2$ by a real Higgs field
$\Sigma_{jk}$
($j,k = 1,2,3$) in the 5-dimensional tensor
representation and whose vacuum expectation value (VEV) has the form
\begin{equation}
 \Sigma_0 = diag(1,1,-2)\sigma
\label{vev}
\end{equation}

Along the path enclosing the straight infinite string the VEV of $\Sigma$
winds by an $SO(3)$ group transformation

\begin{equation}
\Sigma (\theta) = U(\theta)\Sigma_0 U(\theta)
\label{wind}
\end{equation}

interpolating between $Z_2 (=U(2\pi))$ and identity $(=U(0))$, where
$\Delta \theta = 2\pi$ along the path. For instance, we can choose

\begin{equation}
U(\theta) = exp(iT^1 \theta /2)
\end{equation}
where $T^1$ is the generator of  rotations in the 2-3 plane.
 The string carries a topologically stable magnetic flux of the gauge field
$A^1_{\mu}$ associated with the generator $T^1$.
The above  string has the well  known ``Alice'' proprerty that the
$U(1)$-charge
of the point particle gets  flipped when  the later is  transported
around the string. This is due to the fact that generator of $Z_2$
conjugates the $U(1)$-charge.

Let us imagine now that there is an additional Higgs doublet field
$\Phi = (\phi_1, \phi_2)$ in the spinorial representation, which
breaks the gauge symmetry completely.
We  assume that the VEV-s of $\Phi$ and $\Sigma$ are of the same order of
magnitude. Then, the gauge symmetry is broken completely  {\em at one and the
same}
 stage  and
its quotient space is topologically trivial. So naively one does
not    expect formation of any flux carrying (or even global) stable defects in
this model. However this assumption is false: at the classical level
this model contains the {\em topologically} stable semiglobal Alice string.
 The key point is that the most general $SU(2)$-symmetric renormalizable
potential of the fields $\Sigma$ and $\Phi$ is invariant under a larger
global $G_{gl} = SO(3)_{\Sigma} \otimes SU(2)_{\Phi}$ symmetry of independent
global transformations (the index indicates on which field the given
group transformation acts). This is a consequence of the group-theoretical
fact that the only possible (up to a quartic) $SU(2)$-invariant constructed
out of the  one 5-dimensional and one spinorial representation is the
trivial one $(Tr \Sigma^2)(\Phi^{\dag}\Phi)$.

More importantly perhaps, $G_{gl}$ is not  just a symmetry of  the Higgs
potential only. In fact it is the actual global symmetry of the
{\em full} Lagrangian on the arbitrarily fixed $\Phi = constant$ surfaces.
This can be readily seen from the expression of the most general
renormalizable $SU(2)$-gauge invariant Lagrangian of $\Sigma$ and $\Phi$
which has the following form:

\begin{eqnarray}
 L& =& \left[igA^a_{\mu}\Phi^{\dag}  {\sigma^a  \over 2}\partial_{\mu} \Phi  +
hc \right] +
(\partial_{\mu} \Phi )^{\dag}(\partial_{\mu} \Phi ) + {g^2 \over 4}
(A^a_{\mu}A^{a\mu})\Phi^{\dag}\Phi + \nonumber\\
& & tr\left[\partial_{\mu} \Sigma - igA^a_{\mu}[T^a ,\Sigma] \right]^2 - { 1
\over 4}
trF_{\mu \nu}F^{\mu \nu} - V(\Sigma, \Phi )
\label{lagrangian}
\end{eqnarray}

where

\begin{eqnarray}
 V(\Sigma, \Phi ) &=& m^2_1tr\Sigma^2 + h_1(tr\Sigma^2)^2 + h_2tr\Sigma^4 +
m_2tr\Sigma^3 \nonumber\\
 &+& h_3(tr\Sigma^2)(\Phi^{\dag} \Phi ) + m_3(\Phi^{\dag} \Phi ) +
h_4(\Phi^{\dag} \Phi )^2
\label{potential}
\end{eqnarray}

and $T^a$ and $\sigma^a$ ($a=1,2,3$) are the $SU(2)_l$ generators in the
vectorial and spinorial representations respectively, and other indices are
suppressed. Obviously  all  the terms in (\ref{lagrangian}), except
the first one, are invariant under  the independent global transformations:

\begin{equation}
\Sigma\rightarrow U_{\Sigma}^{-1} \Sigma U_{\Sigma} \; ; \; \;
A_{\mu} \rightarrow U_{\Sigma} A_{\mu};
\Phi \rightarrow U_{\Phi} \Phi;
\label{transfo}
\end{equation}

where $U_{\Sigma}=exp[i\theta^aT^a], U_{\Phi}=exp[i\beta^b \sigma^b]$,
and $\theta^a (a=1,2,3)$ and $\beta^b (b=1,2,3)$ are two arbitrary
independent sets of group parameters, and $A_\mu = (A_\mu^1 ,  A_\mu^2 ,
A_\mu^3) $.
 This means that the above
transformation is an exact symmetry of the action on arbitrary
$\Phi=constant$ surfaces (where  the first two terms are zero) and
has an important impact on the vacuum structure.

 The only obvious assumption we make about the structure of the Higgs
potential is that it has a stable global minimum for VEVs of $\Sigma$
and $\Phi$ being nonzero and comparable. It can be checked readily
that the only possible nonzero configurations of the VEV  of $\Sigma$ in
the minimum are those that can be obtained from   (\ref{vev}) by all possible
$U_{\Sigma}$ transformations. Of course the relative orientation of the
VEVs $\Sigma$ and $\Phi$ are arbitrary, which simply reflects the
global (\ref{transfo}) symmetry of the potential. We can fix one of the fields,
say $\Phi$,
 by  a gauge transformation in the form $\Phi =(\phi, 0)$
with $\phi$ real and positive. Then the theory will be minimized
by arbitrary
\begin{equation}
 \Sigma = U_{\Sigma}^{-1} \Sigma_0 U_{\Sigma}
\end{equation}

where $\Sigma_0$ is given in (\ref{vev}). Different group elements $U_{\Sigma}$
bring us to the different points of the continously degenerated vacuum.
These points strictly speaking are not physically equivalent, since
(\ref{transfo})
is not a true symmetry of the theory. However this vacuum misalingment is
rather unusual. {\em Different points of the vacuum (obtained by
different $U_{\Sigma}$) correspond to one and the same pattern of the gauge
symmetry breaking and to  one and the same particle spectrum}.
(Of course the mass eigenstates vary from point to point). This is closely
related to the fact that (\ref{transfo}) is the actual symmetry for $\phi =
constant$.
This implies further that  the misalingment is
not resolved by one loop radiative corrections and the  one loop effective
potential remains degenerate under (\ref{transfo}) (recall that the one-loop
 correction depends only on the particle spectrum \cite{CW}).

 Now let us turn to the topological structure of the vacuum. The global
symmetry (\ref{transfo}) that leaves invariant  the classical (even one loop)
 potential is
spontaneously broken down to $H_{gl}= (U(1) \otimes Z_2)_{\Sigma}$ which
entirely belongs to $SO(3)_{\Sigma}$,  and the vacuum manifold has  a
nontrivial
first homotopy group. So there are topologically stable strings in this
model around which $\Sigma$ winds as in  (\ref{wind}); the corresponding group
transformation that interpolates between identity and $Z_2$ now entirely
belongs to $SO(3)_{\Sigma}$ and is given by (\ref{wind}). The most important
point
about these strings is that they carry  a topologically stable gauge flux.

 Consider again
the closed path that encircles the string at  infinity.
Let us now, as before, fix the gauge with $\Phi = (\phi,0)$ being real and
 positive everywhere.
 Finiteness of the energy forces $\phi$ to be constant along the path.  In
 contrast $\Sigma$ cannot stay constant due to
topological reasons, since it has to wind by  the $U_{\Sigma}$  transformation
that interpolates between $1$ and $Z_2$. We choose this transformation in
the form

\begin{equation}
\Sigma= exp(-iT^1\theta /2) \Sigma_0 exp(iT^1\theta /2)
\end{equation}

where  $0< \theta (x_{\mu}) <2 \pi$ is an continous function of the
spatial point. Whatever the explicit form of the function $\theta$ is,
it should change by $\Delta \theta =2 \pi$ encircling the string.
That is, we have the topological constraint

\begin{equation}
\frac {1}{2\pi} \oint \partial_{\mu} \theta (x) dx^{\mu} = 1
\label{topcons}
\end{equation}

where integration is performed along the same path.
Using  expression (\ref{lagrangian}), we find the following  equation of motion
for
the gauge field $A^1_{\mu}$ at  spatial infinity:

\begin{equation}
A^1_{\mu} = \frac {\partial_{\mu} \theta}{g}(1 + \frac
{\phi^2}{72\sigma^2})^{-1}
\end{equation}

Now performing integration along the same closed path and using  the
topological
constraint (\ref{topcons}) we obtain  the following expression for the gauge
flux:

\begin{equation}
flux = \frac {2 \pi}{g} (1 + \frac {\phi^2}{72\sigma^2})^{-1}
\end{equation}

which is clearly non-integer \cite{semiglobal}. This makes semi-global
strings  perfect candidates for the Aharonov-Bohm type enhancements of the
fermion-string scattering amplitude \cite{AB}.

{\bf Matter Fields}. The physical picture depends sensitively on how the
fermions are
coupled to the scalar fields of the theory, i.e. in their
representation content under the gauge group. Let us discuss the
possible realizations case by case.

\begin{enumerate}
\item {\em Imagine the fermion fields to be triplets under
$SU(2)_l$}, in which case they are coupled to the field
$\Sigma$. This is the conventional case of the Alice  string
discussed in the
literature \cite{alice}. Since both the fermions and the gauge
fields are triplets, they are rotated in the same manner as
they go around the string and thus locally a ``fermionic'' observer will
see no difference (including one loop effects). However, since the
$Z_2$ generator reverses the
signs of `electric' charges, after the full rotation a `particle'
 ends up as an `antiparticle'. Of course, in this case the `photon'
is not massless since $SU(2)_l$ is broken completely; it is just the
lightest gauge boson.

\item {\em Fermions coupled to the doublet $\Phi$}, by say
left-handed fields forming doublets and the right-handed ones singlets
under the gauge group. Now, since $\Phi$ remains constant as it travels
around the string, the same will be true of  the fermionic fields. On
the other hand, the gauge fields, as we said, change (the eigenstates,
not the eigenvalue), and thus, in this case the physical consequences
vary {\em locally} frome one point to another, including the phenomenon
of flavor changing.

\item {\em The fermions decoupled from both $\Sigma$ and $\Phi$}. In
this case, the situation is basically equivalent to the one described
in 2.

\end{enumerate}

{\bf Discussion}
 The simple models we have discussed play a direct role in the theories
of family unification, along the lines discussed by us before
\cite{flavor}.

Among the realistic models the ones that often exhibit nonabelian
accidental degeneracy of the vacuum are sypersymmetric GUTs.
 In these models
the global degeneracy of the vacuum can result from two sources: the group
structure of the Higgs sector (as in nonsupersymmetric case)  and the
supersymmetry. Due to  the usual nonrenormalization theorem,
supersymmetry can support the accidental degeneracy in the vacuum manifold even
if this
degeneracy is not respected by  the gauge  structure of the theory.

We will address those and related issues in detail in a future publication.

\end{document}